\documentclass[12pt]{article}       
\usepackage{amssymb}
\begin{document} 
\begin{center}
          {\large \bf The real part of the elastic scattering amplitude \\ at 
nonzero transferred momenta } 

\vspace{0.5cm}                   
{\bf I.M. Dremin}

\vspace{0.5cm}              
          Lebedev Physical Institute, Moscow 119991, Russia

\end{center}
\begin{abstract}
The model-independent solution of the $s$-channel unitarity condition for the 
imaginary part of the hadronic elastic scattering amplitude outside the 
diffraction peak allows to make conclusions about its real part at nonzero 
transferred momenta. The asymptotical properties of the ratio of the real to 
imaginary part of the amplitude are discussed. In particular, it is  
explicitly shown that the ratio changes its sign at a definite value of the 
transferred momentum. Some comments concerning the present day experimental 
results about the behavior of the differential cross-section of elastic 
scattering outside the diffraction cone are given.
\end{abstract}

The real part of the hadronic elastic scattering amplitudes is measured in
present experiments only at very small transferred momenta in the region
of the Coulomb-nuclear interference. There exist numerous phenomenological 
models which "predict" its behavior at larger values of $\vert t\vert$.
However, moving to new energies we find each time that these attempts are
not very successful. This happened again when TOTEM data for elastic 
$pp$-scattering at the LHC energy $\sqrt s$=7 TeV were published \cite{totem}. 
It is clearly demonstrated by Fig. 2 in Ref. \cite{totem}. All the models 
presented there as well as many newly published ones fail 
to agree with experimental data about behavior of the differential 
cross-section just in the region outside the diffraction peak.

At the same time, it is possible to describe this region in TOTEM data
\cite{dnec} using the simple theoretical explanation based on rigorous 
consequences of the $s$-channel unitarity relation proposed long time ago 
in Refs \cite{anddre, anddre1}. The careful fit to earlier data at rather low 
energies showed also good quantitative agreement with experiment \cite{adg}.

It was shown in Refs \cite{anddre, anddre1} that the unitarity relation outside 
the diffraction peak can be reduced to the integral equation for the imaginary 
part of the amplitude. Its analytic solution was obtained in a model-independent
way. No assumptions have been made other than the validity of experimental 
data in the diffraction peak. It helps get some knowledge about the real part 
of the amplitude at nonzero transferred momenta. 

The $s$-channel unitarity relation is written as
\begin{eqnarray}
{\rm Im}A(p,\theta )= I_2(p,\theta )+F(p,\theta )= \nonumber  \\
\frac {1}{32\pi ^2}\int \int d\theta _1
d\theta _2\frac {\sin \theta _1\sin \theta _2
{\rm Im}A(p,\theta _1){\rm Im}A(p,\theta _2)(1+\rho _1\rho _2)}
{\sqrt {[\cos \theta -\cos (\theta _1+\theta _2)] 
[\cos (\theta _1 -\theta _2) -\cos \theta ]}}+F(p,\theta ).
\label{unit}
\end{eqnarray}
Here $p$ and $\theta $ denote the momentum and the scattering angle in the 
center of mass system. $\rho _i$'s take into account the real parts at the
corresponding angles. 
The region of integration over angles in Eq. (\ref{unit}) is given by the conditions
\begin{equation}
\vert \theta _1 -\theta _2\vert\leq \theta ,       \;\;\;\;\;
\theta \leq \theta _1 +\theta _2 \leq 2\pi -\theta .
\label{integr}
\end{equation}
The integral term represents the two-particle intermediate states of the 
incoming particles. The function $F(p,\theta )$, called following
Ref. \cite{hove} as the overlap function, represents the shadowing contribution 
of the inelastic processes to the elastic scattering amplitude. It determines 
the main structure in the shape of the diffraction peak and is completely 
non-perturbative so that only some phenomenological models pretend to describe 
it. 

The elastic scattering proceeds mostly at small angles. The diffraction peak
has a Gaussian shape in the scattering angles or exponentially decreasing as
the function of the transferred momentum squared  
\begin{equation}\
\frac {d\sigma }{dt}/\left( \frac {d\sigma }{dt}\right )_{t=0}=e^{Bt}\approx 
e^{-Bp^2\theta ^2}.
\label{diff}
\end{equation}
The four-momentum transfer squared is
\begin{equation}
t=-2p^2(1-\cos \theta )\approx -p^2\theta ^2.
\label{trans}
\end{equation}
At large energies the forward scattering amplitude has a small real part as
known from the dispersion relations \cite{drna, blha}. Then the 
elastic scattering in this region labeled by the subscript $d$ can be described 
by the amplitude 
\begin{equation}
A_d(p,\theta )= 4ip^2\sigma _te^{-Bp^2\theta ^2/2}(1-i\rho _d)
\label{ampl}
\end{equation}
with a proper optical theorem normalization to the total cross-section 
$\sigma _t$ in the forward direction and small correction due to the real part.

Now, let us consider the integral term $I_2$ outside the diffraction peak.
Because of the sharp fall-off of the amplitude (\ref{ampl}) with angle, the 
principal contribution to the integral arises from a narrow region around the
line $\theta _1 +\theta _2 \approx \theta $. Therefore one of the amplitudes
should be inserted at small angles within the cone while another one is kept 
at angles outside it. Thus inserting 
Eq. (\ref{ampl}) for one of the amplitudes in $I_2$ and integrating over one 
of the angles the inhomogeneous linear integral equation is obtained:
\begin{equation}
{\rm Im}A(p,\theta )=\frac {p\sigma _t}{4\pi \sqrt {2\pi B}}\int _{-\infty }
^{+\infty }d\theta _1 e^{-Bp^2(\theta -\theta _1)^2/2} (1+\rho _d\rho _l)
{\rm Im}A(p,\theta _1)+F(p,\theta ).
\label{linear}
\end{equation}
It can be solved analytically (for more details see \cite{anddre, anddre1})
with the assumptions that the role of the overlap function $F(p,\theta )$ is 
negligible outside the diffraction cone\footnote {The results of the papers 
\cite{dnec, ads} give strong support to this assumption.} and the real parts
may be replaced by their average values in the diffraction peak $\rho _d$ and
outside it $\rho _l$, correspondingly. Let us stress once more that the Gaussian
shape (\ref{ampl}) of the amplitude has been only used at rather small angles
in accordance with experimental data.

Using the Fourier-Bessel direct and inverse transformations 
one gets the analytic solution
\begin{equation}
{\rm Im} A(p,\theta )=C_0(p)e^{-\sqrt 
{2B\ln \frac {4\pi B}{\sigma _t(1+\rho _d\rho _l)}}p\theta }+
 \sum _{n=1}^{\infty }C_n(p)
e^{-({\rm Re }b_n(p))p\theta } \cos (\vert {\rm Im }b_n(p)\vert p\theta-\phi _n)
\label{sol}
\end{equation}

\begin{equation}
b_n\approx \sqrt {2\pi B\vert n\vert}(1+i{\rm sign }n) \;\;\;\;\;\;\; n=\pm 1, \pm 2, ...
\end{equation}

This shape has been obtained from contributions due to the pole on the real axis
and a set of the pairs of complex conjugated poles in the direct  
Fourier-Bessel transform of the equation. Correspondingly, the solution
contains the exponentially decreasing with $\theta $ (or 
$\sqrt {\vert t \vert }$) term (Orear regime!) with imposed on it 
damped oscillations. 

Note that the solution predicts the dependence on $p\theta \approx 
\sqrt {\vert t\vert }$ but not the dependence on the collision energy!
There are no zeros on the $t$-axis unless the amplitudes of oscillations
$C_n(p)$ become extremely large.

Namely this expression was successfully used in Ref. \cite{dnec} to fit 
the elastic scattering 
differential cross-section at 7 TeV outside the diffraction cone (in the Orear 
regime region). The first (Orear) term is exponentially decreasing with $\theta $ 
(or $\sqrt {\vert t \vert }$) and the next terms demonstrate the damped 
($n\geq 1$) oscillations which are in charge of the dip-maximum structure near 
the diffraction cone. The values of the slope $B$ and total cross-section 
$\sigma _t$ determine mostly the shape of the elastic differential cross-section 
in the Orear region which is placed between the Gaussian diffraction peak and 
the power-like decreasing regime of the large angle parton 
scattering. The value of $4\pi B/\sigma _t$ is so close to 1 at 7 TeV that 
the first term is very sensitive to the ratio $\rho _l$ outside the
diffraction peak.  Thus it became possible for the first time to estimate 
$\rho _l$ from fits of experimental data and it happened to be quite large
($\rho _l\approx -2$). Concerning the ratio $\rho _d $ it was chosen  
as prescribed by the dispersion relations for its value at $t=0$ 
\cite{drna, blha} ($\rho _d\approx 0.14$).

Our main concern in this paper is, however, not in the fit of experimental data
at present energies as was done in Ref. \cite{dnec} but in considering possible 
asymptotic conclusions from the solution (\ref{sol}). The main role is played
there by the first term and we discuss it. According to general belief stated
as early as in the 1960s (e.g. see Refs \cite{froi, mart, chou, chen}) $\sigma _t$ 
approaches infinity as
\begin{equation}
\sigma _t=2\pi R^2+O(\ln s); \;\;\;\; R=R_0\ln s; \;\;\;\; R_0={\rm const},
\label{cst}
\end{equation}
bounded from above \cite{froi} by
\begin{equation}
\sigma _t\leq \frac {\pi }{m^2_{\pi }}\ln ^2s,
\label{frois}
\end{equation}
where $m_{\pi }$ is the pion mass.
 
The width of the diffraction peak $B^{-1}(s)$ should shrink if
\begin{equation}
B(s) \propto R\propto \ln s \;\;\;\;\;  {\rm or}  \;\;\;\;\; 
\propto R^2\propto \ln^2s \;\;\;\; \cite{kino}.
\label{wid}
\end{equation}
The ratio of the real part to the imaginary part of the amplitude in the 
forward direction must vanish asymptotically as
\begin{equation}
\rho (s,0)=\frac {\pi }{\ln s}+O(\ln ^{-2}s).
\label{rho0}
\end{equation}
These three parameters define the exponent in the first leading term in 
Eq. (\ref{sol}). If $B(s)\propto \ln^2s$ asymptotically, then the logarithm in 
this (Orear)  term tends to some constant. One can write it as
\begin{equation}
{\rm Im}A_o(s,t)=C_0(s)\exp (-\sqrt {\tau })=C_0(s)f(\tau ),
\label{ore}
\end{equation}
where $\tau =(t/t_0)\ln^2s$.

The amplitudes of this type are well studied \cite{aube, mar1}. They lead to the 
asymptotical geometrical scaling. The hypothesis of geometrical scaling was
first proposed in Ref. \cite{ddd} for description of hadron collisions at finite 
energies in a wide range of transferred momenta. 

It is important 
that the ratio of real and imaginary parts of the amplitude can be calculated 
\cite{mar1} at nonzero transferred momenta $t$ as
\begin{equation}
\rho =\frac{\pi }{\ln s} \left [1+\frac {\tau (df(\tau )/d\tau )}{f(\tau )}
\right ].
\label{rhotau}
\end{equation}

Now one is able to do this explicitly since the imaginary part of the fixed 
$t<0$ scattering amplitude is known from Eq. (\ref{sol}) at finite energies. With 
the assumption $B(s)\propto \ln^2s$ one can use Eq. (\ref{ore}). The result is
\begin{equation}
\rho (s,t)=\frac {\pi}{\ln s}\left [1-\frac {a\sqrt {\vert t\vert }}{2}\right ]
=\rho (s,0)\left [1-\frac {a\sqrt {\vert t\vert }}{2}\right ],
\label{rhot}
\end{equation}
where
\begin{equation}
a=\sqrt {2B\ln \frac {4\pi B}{\sigma _t(1+\rho _d\rho _l)}}.
\label{a}
\end{equation}
We note that $\rho $ passes through zero and changes sign at 
$\vert t\vert =4/a^2$. This agrees with the general theorem on the change of 
sign of the real part of the high-energy scattering amplitude which has been 
proven first in Ref. \cite{mar2}.

Let us try to come back to the present day energies. 
Strictly speaking, the formulae (\ref{rhotau}), (\ref{rhot}) can only be 
applied at asymptotic energies at the assumption of the $\ln ^2s$ behavior
of the total cross-section and for small enough $\vert t\vert $. Nevertheless,
it is tempting to get some naive estimates at present values of $s$ and $t$
ignoring these precautions in attempts to understand where do we stand now.

Naively, one can insert Eq. (\ref{rhot}) in the expression for the elastic
differential cross-section to get
\begin{equation}
\frac {d\sigma (s)}{dt}=\frac {1}{16\pi s^2}({\rm Im}A)^2(1+\rho ^2)
=\frac {1}{16\pi s^2}({\rm Im}A(s,t))^2[1+\rho ^2(s,0) (1-
0.5a\sqrt {\vert t\vert })^2].
\label{naiv}
\end{equation}
The slope of the differential cross-section in the Orear region is given by 
$2a$. To fit it one should get $a\approx 2.73$ GeV$^{-1}$ at ISR energies and 
$a\approx 3.25$ GeV$^{-1}$ at LHC. The contribution of the second term due to 
the real part of the amplitude in (\ref{naiv}) is negligibly small because
$\rho (s,t)$ changes in the range between 0 and $-\rho (s,0)\approx -0.14$ 
in the Orear region ($0.3<\vert t\vert <1.5$ GeV$^2$ at LHC). Its average 
value there is surely much smaller (in the absolute value) than what is 
needed ($\rho _l\approx -2$ in Ref. \cite{dnec}) to get the proper slope at 
7 TeV. Such a large difference between estimates of the contribution of
the real part of the elastic amplitude to the differential cross-section
in experiment and according to (\ref{rhot}) poses a problem. No agreement in
absolute values of $\rho $ is found in the Orear region. Now one can claim 
only the qualitative agreement with the awaited change of sign \cite{mar2} 
of $\rho _(s,t)$ compared with $\rho (s,0)$. Actually, this demonstrates that 
we are still in preasymptotical regime and the geometrical scaling used for 
derivation of Eqs (\ref{rhotau}), (\ref{rhot}) does not hold at present 
energies.

The most important parameter defining the slope is the value of the ratio 
$4\pi B/\sigma _t$ in Eq. (\ref{sol}). According to experimental data it 
is close to 1 in a wide energy range. Actually, using the Table II in 
Ref.\cite{chao} we estimate that it is about 1 at $\sqrt s$ = 4 GeV, increases 
almost to 1.5 at ISR energies and then again drops near 1 at 7 TeV. 
The slope and the total cross-section increase almost at the same 
rate in the ISR energy range (see, e.g., Fig. 1 in Ref. \cite{bddd}). 
They are however proportional approximately to $\ln s$ but not to $\ln ^2s$.
Thus the geometrical scaling is not precise even there.
At 7 TeV the ratio $4\pi B/\sigma _t$ becomes almost equal to 1 because 
cross-section has increased faster than the slope. The preasymptotical regime 
of the power-like growth of the total cross-section is at work at present 
energies while the cone shrinkage is less active. Thus the fit became very 
sensitive to the value of $\rho _l$. The ratio 
$4\pi B/\sigma _t$ in the combination with values of $\rho _d$ and $\rho _l$ 
at different angles determines the slope $2a$ of the differential 
cross-sections in the Orear regions at any $s$.

The position of zero in (\ref{rhot}) changes from $\vert t\vert $=0.54 GeV$^2$ 
at ISR to 0.36 GeV$^2$ at LHC, i.e. it is at the edge of the diffraction peak.
The solution (\ref{sol}) is at the limit of its applicability there.
Asymptotically this zero moves to $\vert t\vert $=0 as the width of the
diffraction peak $B^{-1}(s)$ shrinks unless the role of the logarithm in 
(\ref{a}) becomes important. The average of $\rho (s,t)$ in the diffraction cone 
denoted by $\rho _d$ is very close to $\rho (s,0)$ if one uses (\ref{rhot}).
Thus its replacement by $\rho (s,0)$ in Ref. \cite{dnec} is quite reasonable.

According to (\ref{rhot}) the value of $\rho (s,0)$ decreases logarithmically 
with energy. However, it is yet somewhat higher (about 0.177) than estimates 
from dispersion relations (0.14 in Refs \cite{drna, blha}) even at 7 TeV 
and strongly overshoots them at ISR where $\pi /\ln s \approx 0.37$.
No logarithmic decrease is seen in these estimates. Moreover, the value 0.14
can only be reached according to (\ref{rhot}) at the energy 75 TeV. Probably,
at energies higher than 75 TeV the first signs of approach to asymptotics will 
become visible. 

What is more exciting, some new effects can appear at higher energies
according to the formula (\ref{sol}). The experimentally observed decrease of 
the parameter $4\pi B/\sigma _t$ with energy leads to interesting predictions 
if the whole argument of the logarithm in 
(\ref{sol}) becomes less than 1. Then the pole in the Fourier-Bessel transform 
moves from the real axis to the imaginary one. The Orear region disappears and
a new regime with nondamped oscillations at $\vert t\vert $ outside the 
diffraction peak starts to play a role. Such possibility was considered 
using a definite model with multiple Pomeron exchanges in Regge-approach 
a long time ago \cite{adya}. There is a close correspondence with the present
approach because the solution (\ref{sol}) can be treated as multiple
iterations of the diffraction cone regime (or of the overlap function 
$F(p,\theta )$ in the inhomogeneous equation (\ref{linear})
as it is done in Ref \cite{anddre1}).

The preasymptotical nature of presently observed effects is also seen from small 
value of the ratio of the elastic to total cross-section which increased from 
about 0.2 at ISR to 0.25 at LHC. Its expected asymptotic value 0.5 would 
correspond to the black disk limit. 

We conclude that even though the qualitative trends of experimental data may be 
considered as rather satisfactory ones for theoretical prejudices, we are sitll 
pretty far from asymptotics even at the LHC energies.
 
\medskip

{\bf Acknowledgement}

This work was supported by the RFBR grant 12-02-91504-CERN-a and by the 
RAN-CERN program.

\end{document}